\documentclass{article}
\usepackage{spconf,amsmath,graphicx,hyperref, booktabs}
\usepackage{multirow, amssymb, comment}

\usepackage{soul, color}
\usepackage{standalone}

\title{Incremental learning for audio classification with Hebbian Deep Neural Networks}

\name{Riccardo Casciotti$^1$, Francesco De Santis$^2$, Alberto Antonietti$^2$, Annamaria Mesaros$^1$ \thanks{ 
We acknowledge CSC Finland for awarding this project access to the LUMI supercomputer, owned by the EuroHPC Joint Undertaking, hosted by CSC (Finland) and the LUMI consortium through Jane and Aatos Erkko Foundation, grant number 230048, "Continual learning of sounds with deep neural networks".}
}

\address{$^1$ Signal Processing Research Centre, \textit{Tampere University}, Tampere, Finland \\
$^2$ Department of Electronics, Information and Bioengineering (DEIB), Politecnico di Milano, Italy
\\
\{riccardo.casciotti, annamaria.mesaros\}@tuni.fi, \{alberto.antonietti, francesco.desantis\}@polimi.it
}

\begin{document}
\ninept

\maketitle

\begin{abstract}
The ability of humans for lifelong learning is an inspiration for deep learning methods and in particular for continual learning. In this work, we apply Hebbian learning, a biologically inspired learning process, to sound classification. We propose a kernel plasticity approach that selectively modulates network kernels during incremental learning, acting on selected kernels to learn new information and on others to retain previous knowledge. Using the ESC-50 dataset, the proposed method achieves 76.3\% overall accuracy over five incremental steps, outperforming a baseline without kernel plasticity (68.7\%) and demonstrating significantly greater stability across tasks.

\end{abstract}

\begin{keywords}
Audio classification, Incremental learning, Catastrophic forgetting, Hebbian learning 
\end{keywords}
\section{Introduction}
\label{sec:intro}
Continual learning is the behavior of artificial intelligence models to incrementally acquire new information and learn new patterns, showing robustness and resistance in terms of data distribution shifting and task change \cite{ven_continual_2025}. By default, deep learning models suffer from Catastrophic Forgetting, defined as the abrupt forgetting of previously learned patterns due to the overwriting of the internal representations whenever the model encounters a new data distribution or a new task \cite{ven_continual_2025}. 

The continual learning problem is characterized by a trade-off between learning plasticity and memory retention. It has been shown that incrementally learning new information reduces learning plasticity as the model absorbs new patterns, and that it is crucial for the model to know what to keep from previous patterns and what to forget, stimulating knowledge transfer \cite{ven_continual_2025}. 
Approaches to catastrophic forgetting fall into two categories. Manual strategies include replaying past data \cite{schaul_prioritized_2016}, constraining updates through selective plasticity methods such as Elastic Weight Consolidation \cite{kirkpatrick_overcoming_2017} and Synaptic Intelligence \cite{zenke_continual_2017}, freezing or expanding subnetworks as in Progressive Neural Networks \cite{rusu_progressive_2022} or PathNet \cite{fernando_pathnet_2017}, and regularization methods like Learning without Forgetting \cite{li_learning_2017}. Other work employs gating and attention to reduce interference \cite{masse_alleviating_2018}. A more recent trend is to replace hand-crafted heuristics with meta-learning: Model-Agnostic Meta-Learning (MAML) learns initializations that rapidly adapt to new tasks \cite{finn_model-agnostic_2017}, and Online-aware Meta-Learning (OML) meta-learns representations that remain stable and mitigate forgetting \cite{javed_meta-learning_2019}. Another approach involves acting on the loss function to mitigate catastrophic forgetting \cite{mulimani_closer_2025}. 
All these solutions for coping with catastrophic forgetting were implemented in the context of the backpropagation algorithm. Here, we propose building a different solution, as explained in the following.

Continual learning is inspired by a similar skill humans have perfected with evolution: Lifelong Learning, i.e., the ability to continuously learning new information throughout a person's life. Humans suffer from forgetting in a much lighter manner compared to AI models, and are easily able to recall past information and experiences \cite{parisi_continual_2019}. Inspired by this, 
we explore a biologically plausible \cite{j_schwartz_principles_1991} solution for continual learning based on Hebbian learning. 
\begin{figure*}[!t]
  \centering
  \vspace{-6pt}
  \includegraphics[width=.8\linewidth]{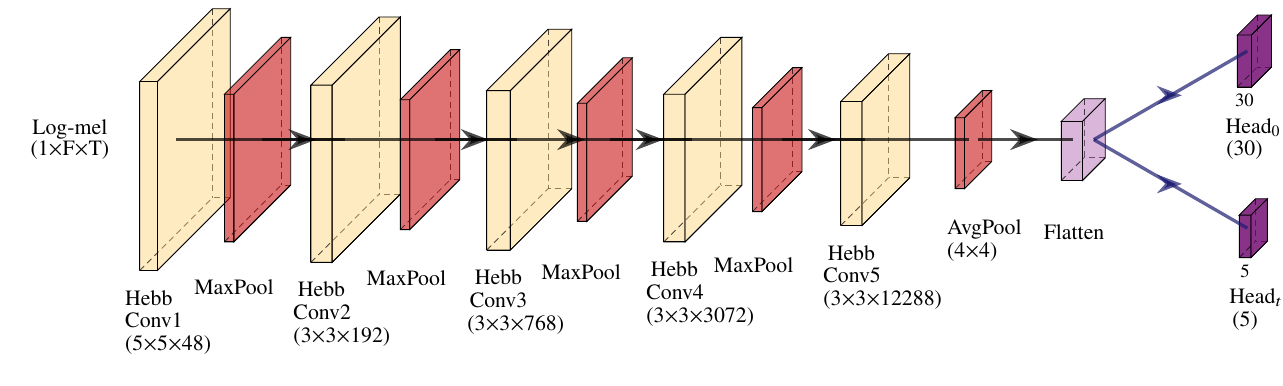}
  
  \vspace{-12pt}
  \caption{Model architecture: Hebbian convolutional layers and task-dependent classifier heads. }
  \label{fig:block-diagram}
    \vspace{-8pt}
\end{figure*}

Hebbian learning can be summarized with the following quote: "neurons that fire together wire together" \cite{munakata_hebbian_2004}. While artificial neural networks typically learn through the backpropagation algorithm, Hebbian learning is much less powerful in learning patterns due to the lack of a feedback signal carrying an error, and the learning being based solely on the correlation between the data points. A solution to this issue is the SoftHebb algorithm, on which we base our approach. SoftHebb modifies the classical Hebbian learning approach by intervening on inter-layer activations and adopting a dynamic learning rate during training.

We perform a learning rate modulation during training, with the objective of overcoming catastrophic forgetting. To the best of our knowledge, this is the first work combining Hebbian learning with incremental learning. We use a task-incremental learning scenario characterized by task-specific classification heads, similar to \cite{vedaldi_adversarial_2020}, with focus on the learning process in the Hebbian network rather than the classifier itself. 

The contributions of our work are: (1) we investigate the use of Hebbian networks and Hebbian learning, a bio-plausible and unsupervised learning architecture and process, for audio classification; (2) we propose a kernel plasticity approach built on learning rate modulation to mitigate catastrophic forgetting of the network during incremental learning. (3) we analyze the forgetting process during the learning stages using specific metrics that evaluate the ability of the network to retain previous tasks while learning new ones.
This paper is organized as follows: Section \ref{sec:method} describes the Hebbian learning and the proposed kernel plasticity method used in conjunction with a multi-head classifier for task-incremental learning. Section \ref{sec:experim-setup} explains the dataset, training and test procedure, and evaluation metrics used to quantize the forgetting process\footnote{Code available at https://github.com/RiccardoCasciotti/Hebbian-TIL}. Section \ref{sec:results} presents the experimental results and comparisons to different learning alternatives. Section \ref{sec:concl} presents conclusions and future work.

\section{Hebbian learning for audio classification}
\label{sec:method}
Hebbian learning is a principle describing associative learning, in which neurons strengthen their synaptic connections when they are active simultaneously. Hebbian learning uses only the correlation between the samples to learn new information, thus not needing feedback information.
The model used in this work is based on the SoftHebb architecture \cite{journe_hebbian_2023}, which introduced a new Hebbian learning paradigm  by leveraging a Bayesian interpretation to the outputs of each Hebbian layer. The work in \cite{journe_hebbian_2023} also introduced an anti-Hebbian plasticity that limits exploding weights, and an adaptive learning rate based on weight convergence during training. 
The model was used 
to perform image classification using different datasets such as CIFAR-10, CIFAR-100 \cite{krizhevsky_learning_2009}, and others,  reaching performances comparable to backpropagation-based models.

In this work, we use SoftHebb to solve a task-incremental learning scenario; we sequentially train the model on new tasks, and the task label is known both during training and inference \cite{ven_three_2019}.

The proposed architecture is presented in Fig.~\ref{fig:block-diagram} and consists of a feature extractor, which has five convolutional layers that use Hebbian learning, and a classification layer that predicts the class labels. The classification layer is a fully-connected layer that we train using backpropagation in order to perform classification. 
The system is trained in a "consecutive" fashion \cite{journe_hebbian_2023}: the feature extractor is trained first through a forward pass using the SoftHebb algorithm (unsupervised), in only one epoch. The feature extractor is then frozen, and the classifier head is trained using backpropagation.

The continual learning solution implemented in this work consists of two separate parts: a solution applied to the feature extractor and a solution applied to the classifier head of the model.  Through this double approach, we aim to protect both the weights that store the representation embedding of the information (feature extractor) and the weights that learn how to interpret those embeddings (classifier head) \cite{wang_comprehensive_2024}.

\subsection{Kernel Plasticity}
The kernel plasticity solution applied to the feature extractor takes inspiration from the natural neurotransmitters present in the brain, such as dopamine, used to cement newly learned information and preserve it in time \cite{magotra_neuromodulated_2021}. 
We use neuro-modulation of the learning process through learning rate updates on specific kernels. 

Kernel selection is done in two steps: 
(1) check the average weight update for the kernels during training on a task; (2) rank the kernels based on their activation value during training on a task.
The first step aims to understand how the kernels change during training on a task by keeping track of the weights every \textit{i} batches and computing how they have changed with respect to the previous step by calculating the difference. By cumulatively summing all the cells of this weight difference matrix, a single weight change value is computed and associated to each kernel.

For a kernel \( j \), let the weight at the beginning of the \( i \)th interval be \( w_j^{(i-1)} \) and at the end be \( w_j^{(i)} \).  The weight change during interval \( i \) is given by:
\vspace{-2pt}
\begin{equation}
\Delta w_j^{(i)} = w_j^{(i)} - w_j^{(i-1)}.
\end{equation}
The average weight change over \( N \) intervals can be defined using the cumulative change:
\begin{equation}
\overline{\Delta w_j} = \frac{1}{N} \sum_{i=1}^{N} |\Delta w_j^{(i)}|.
\end{equation}
\vspace{-2pt}
At the end of this process, a vector containing the average weight change per every kernel in every layer is stored in the model.

The second step calculates a cumulative sum of the activation values for each kernel every \textit{i} batches of data samples. After the training phase is completed, a final activation value is calculated and associated with each kernel. Kernels are sorted based on this value, having the kernels with the highest value at the top of the list. Finally, we consider only the \textit{top k} kernels in this list as important for the just seen task and store them permanently. 
Let the activation value of kernel \( j \) at interval \( i \) be \( a_j^{(i)} \) (an average or sum over spatial dimensions). The cumulative activation for kernel \( j \) is:
\begin{equation}
A_j = \sum_{i=1}^{N} a_j^{(i)}.
\end{equation}
The kernels are ranked based on \( A_j \) and the indices corresponding to the top \( K \) kernels are stored:
\[
\mathcal{K} = \{ j : A_j \text{ is among the top } K \text{ values} \}.
\]

 The neuro-modulation process is applied using the \textit{average kernel weight update} and \textit{top k} kernels per task, as follows, executed during training on a new task: (1) check if the incoming weight update surpasses the average kernel weight change stored in the tensor for previous training sessions; (2) check if the current kernel is among the \textit{top k} kernels through the stored tensor.
If both conditions are met, the incoming weight update is increased on kernels that are not among the \textit{top k} and decreased among the \textit{top k} kernels that break the threshold of the previously stored average weight update. So, for an incoming weight update \( \Delta w_j^{\text{new}} \), first a global modulation condition is defined that checks whether at least one kernel within the \textit{top k} set $K$ surpasses its average weight-update threshold:
\begin{equation}
\text{condition} = \exists\, k \in K : \|\Delta w_k^{new}\| > \Delta w_k
\label{eq:global_condition}
\end{equation}
then, the neuro-modulation of kernel weight update is:
\begin{equation}
\Delta w_j^{mod} = 
\begin{cases}
\alpha\,\Delta w_j^{new}, & \text{if (cond=True) and } j \notin K, \\[6pt]
\beta\,\Delta w_j^{new}, & \text{if } j \in K \text{ and } \|\Delta w_j^{new}\| > \Delta w_j, \\[6pt]
\Delta w_j^{new}, & \text{otherwise},
\end{cases}
\label{eq:neuromodulation_rule}
\end{equation}

\noindent where $\alpha > 1$ enhances plasticity for less-critical kernels, and $0 < \beta < 1$ protects the stability of important kernels. Thus, the increased plasticity for kernels not included in the \textit{top k} set ($j \notin K$) is triggered \emph{only} when at least one kernel from the \textit{top k} kernels surpasses its average threshold, signaling active learning in important kernels.
\subsection{Multi-head classifier}
\label{sub:multi_head}

In line with the task-incremental learning definition, the task label is available to the model during both training and inference. At the training stage, when the model encounters a new task \(T_{\text{new}}\), it proceeds as follows:

\noindent (1) \textbf{Resume the feature extractor:} the feature extractor \(\phi\) is used, with the weights learned so far; 

\noindent (2) \textbf{Instantiate a new head:} A new head \(H_{\text{new}}\) is created. At the beginning of training on task \(T_{\text{new}}\), the network output is given by:
  \begin{equation}
  y = H_{\text{new}}\bigl(\phi(x)\bigr), \quad \text{with } x \in \mathbf{R}^{d}.
  \end{equation}
\noindent (3) \textbf{Train the feature extractor, then train the classifier head and save the head:} After training, the new head \(H_{\text{new}}\) is stored:
  $\mathcal{H} = \mathcal{H} \cup \{H_{\text{new}}\}$.  
During inference, the model will select the head corresponding to the task label by extracting it from the stored vector. The final prediction for an input \(x\) is obtained using the selected head:
\vspace{-2pt}
\begin{equation}  
y = H_{t^*}\bigl(\phi(x)\bigr)
\end{equation}
\vspace{-20pt}
\section{Experimental setup and evaluation}
\label{sec:experim-setup} 
\subsection{Model training procedure}
For the experiments in this work we use the ESC-50 dataset \cite{piczak_esc_2015}, a labeled collection of 2000 environmental audio recordings  routinely used for benchmarking methods in environmental sound classification. The dataset consists of 5-second-long recordings organized into 50 semantical classes with 40 examples per class, organized into 5 folds.
During training, we use one fold for testing, one fold for validation, and the remaining three folds for training.

The incremental learning process is organized into five separate tasks, which the model has to train on incrementally. The first task consists of 30 classes, while the remaining 4 tasks comprise 5 classes each. The tasks are generated by randomly selecting a subset of classes of the dataset, with disjoint subsets of classes per task.
To learn a new task, the model uses a new classifier head, and always keeps the same feature extractor. The consecutive training process is the same in each incremental task: the feature extractor is trained first, with a frozen classifier head, after which the feature extractor is frozen and the classifier head is trained using backpropagation for 50 epochs (determined through validation experiments). When the training for the current task is completed, the task-specific classifier head is stored to be used at test time. 

The feature extractor of the model selected for the experiments has 5 Hebbian convolutional layers,  

as shown in Figure \ref{fig:block-diagram}. Each convolutional layer is characterized by a Triangle activation function and is followed by a max pooling layer, except for the last convolutional which uses an average pooling layer. Additionally, every layer uses batch normalization. We selected the hyperparameters for the layers based on grid search and empirical testing, and taking into account the study done in \textit{SoftHebb} \cite{journe_hebbian_2023}, where a Triangle activation function and the max pooling layers showed to yield better results than the more classical ReLu \cite{dubey_activation_2022} activation function and other pooling solutions.
The incremental learning process is controlled by a number of hyperparameters selected 
using the validation set. The fraction of top kernels \textit{top k} we protect from overwriting is 0.6; the learning rate modifiers for plastic vs important kernels $\alpha$ and $\beta$ 
are 0.15 and 0.9, respectively; the interval (in batches) for tracking kernel updates 
is 5. The range for \textit {k}, and learning rate modifiers 
is [0,1], while the update tracking interval 
ranges from [0, \textit{max number of batches}].

At testing stage, the classifier head corresponding to the task label is used together with the feature extractor. 
After each learned task, the model is tested using data from all tasks learned so far.
\begin{table*}[t]
\centering
\caption{Classification accuracy comparison on ESC-50. For task-incremental learning (TIL), the first value indicates accuracy over all tasks learned so far,  with accuracy over previous tasks, and accuracy over the last learned task in parentheses.}

\begin{tabular}{l|c|ccccc|c}
\toprule
Method & KP & Task 0 & Task 1 & Task 2 & Task 3 & Task 4 & Overall \\
\midrule 

EWC Baseline & - & 9.5 & 54.5 & 63.5 & 82.5 & 70.5 & 33 \\
\midrule
\multirow{2}{4em}{TIL (proposed)}
& --        & \textbf{60.4 }(-, 60.4) & 70.9 (58.1, 83.7) & 72.7 (67.4, 83.4) & 71.2 (67.2, 83.3) & 68.7 (65.1, 83.0) & 68.7 \\
& \checkmark & 60.0 (-, 60.0) & \textbf{71.4} (58.0, 84.7) & \textbf{74.6} (70.5, 82.6) & \textbf{75.8} (73.7, 82.3) & \textbf{76.3} (75.0, 81.0) & \textbf{76.3} \\
\midrule
\multirow{2}{4em}{Joint learning}
& --        & 60.4 & 57.9 & 57.4 & 57.2 & 58.4 & 58.4 \\
& \checkmark & 60.0 & 58.5 & 56.8 & 54.9 & 54.7 & 54.7 \\
\midrule
Common head & -- & -- & -- & -- & -- & -- & 53.3 \\
\bottomrule
\end{tabular}
\vspace{-20pt}
\label{tab:acc}
\end{table*}

\begin{table}[t]
\centering
\caption{Metrics evaluating the incremental learning on ESC-50}.
\vspace{2pt}
\begin{tabular}{l|c|cccc}
\toprule
Metric & KP & Task 1 & Task 2 & Task 3 & Task 4 \\
\midrule
\multirow{2}{*}{BWT}
& --        & -2.33 & -4.67 & -8.64 & -12.63 \\
& \checkmark & -1.98 & -1.82 & -2.11 &  -2.36 \\
\midrule
\multirow{2}{*}{IM}
& --        & -25.85 & -25.91 & -26.11 & -24.61 \\
& \checkmark & -26.22 & -25.83 & -27.36 & -26.33 \\
\midrule
\multirow{2}{*}{FM}
& --        & 2.33 & 1.15 & 1.22 & 1.04 \\
& \checkmark & 1.98 & 0.88 & 0.90 & 0.56 \\
\bottomrule
\end{tabular}
\vspace{-8pt}
\label{tab:metrics}
\end{table}
\subsection{Evaluation metrics}
The performance of the model is evaluated using classification accuracy on the test set. For each experiment, we report the average performance over 10 runs in which the classes are assigned to the 5 incremental tasks randomly. 
We also use a set of metrics to measure the effects of incremental learning. These metrics characterize the forgetting experienced by the model, and allow us to evaluate the knowledge transfer, as well as the memory plasticity and the memory stability of the network across tasks.
\textbf{Forgetting Measure} (
\cite{lopez-paz_gradient_2017} (FM) directly quantifies catastrophic forgetting on previous tasks when learning a new one, defined as:
\vspace{-2pt}
\begin{equation}
\text{FM}_{k} = \frac{1}{k-1} \sum_{j=1}^{k-1} f_{j,k}\label{eq:7} \\
\end{equation}
\vspace{-2pt}
\noindent where 
$f_{j,k} = \max_{i \in \{1,\dots,k-1\}} a_{i,j} - a_{k,j}, \forall j < k $ 
and $a_{i,j}$ is the accuracy on task $i$ after learning task $j$.
\textbf{Backward Transfer} 
\cite{ferrari_riemannian_2018} (BWT) measures the interference of new tasks on previous ones and is defined as:
\vspace{-2pt}
\begin{equation}
\text{BWT}_{k} = \frac{1}{k-1} \sum_{j=1}^{k-1} a_{k,j} - a_{j,j} \label{eq:8} \\
\end{equation}
\vspace{-2pt}

\textbf{Intransigence Measure} 
\cite{ferrari_riemannian_2018}(IM) measures the difference in performance between the model in the incremental learning scenario and the joint model:
\vspace{-2pt}
\begin{equation}
\text{IM}_{k} = a^{*}_{k} - a_{k,k} \label{eq:9}
\end{equation}
\vspace{-2pt}
\noindent where $a^{*}_{k}$ is the accuracy of a reference model trained jointly on task $k$, and $a_{k,k}$ is the accuracy on task $k$ right after training.
\section{Results and discussion}
\label{sec:results}
Table \ref{tab:acc} shows the classification accuracy of different learning variants after each incremental stage. We compare the proposed method with a system that does not use kernel plasticity (KP) in the training, but uses the multi-head Hebbian learning setup. We also provide a EWC-based \cite{kirkpatrick_overcoming_2017} baseline system. We also compare with systems that violate the incremental learning setting, i.e., have access to the data of the previous tasks. For this, we evaluate the same architecture but jointly-trained, i.e., trained directly on the 35, 40, 45, and 50 classes, respectively, at each stage. Finally, we compare with a model that uses the feature extractor trained incrementally in TIL with KP, but a new head trained on all 50 classes at the end of the TIL.

\begin{figure}
    \centering
    \includegraphics[width=1\linewidth]{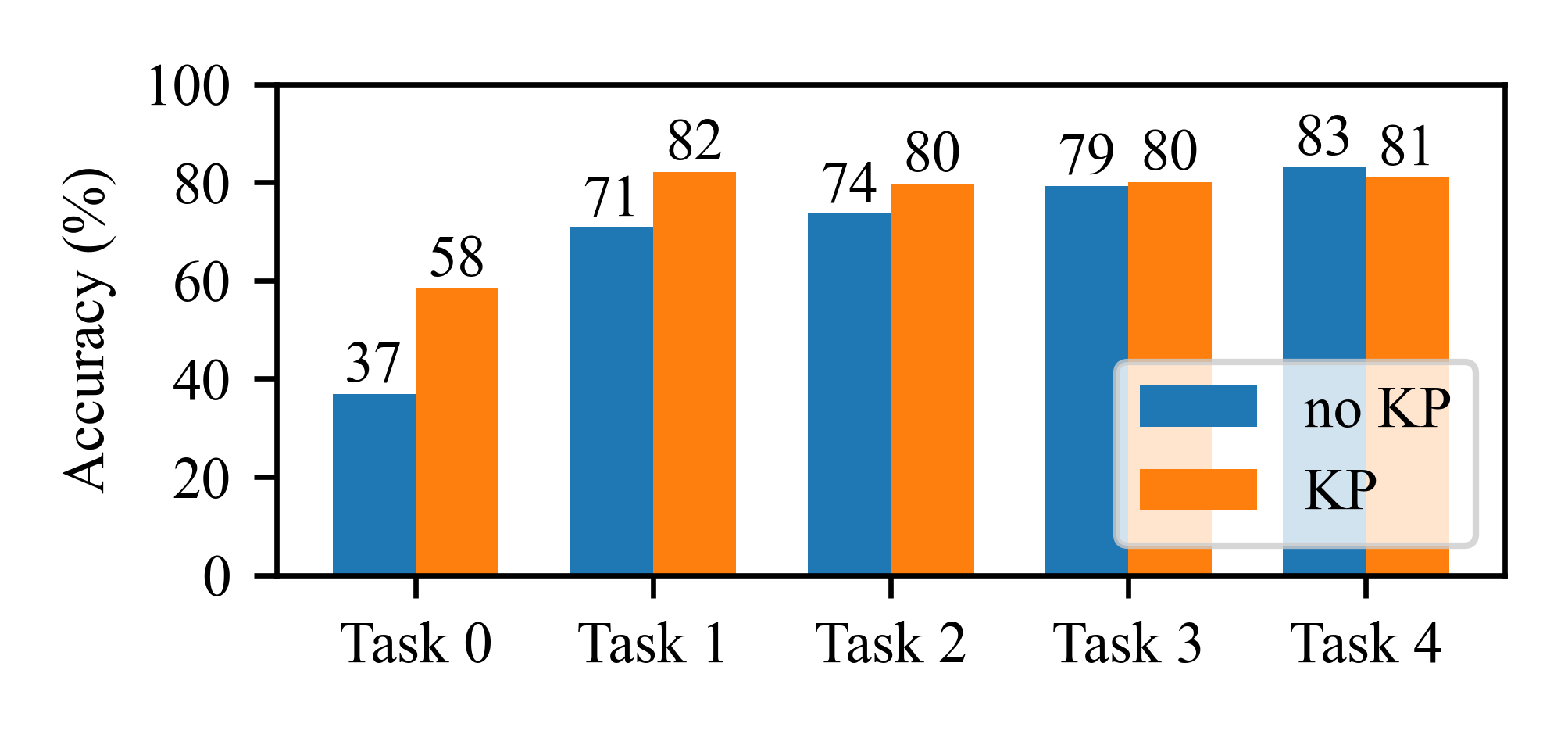}
    \vspace{-24pt}
    \caption{Comparison of the task-wise accuracy between using or not using KP in the incremental learning on ESC-50.}
    \vspace{-10pt}
    \label{fig:barplot}
\end{figure}

Trained jointly on all the 50 classes, the system has an accuracy of 58.4\% (average accuracy over 10 runs). 
The accuracy is significantly lower than for other approaches benchmarked using the same data \cite{gong_ast_2021}; however, here we deal with a completely different learning paradigm, hence a direct comparison is somewhat unfair. As an incremental learning approach, the proposed method outperforms the EWC \cite{kirkpatrick_overcoming_2017} baseline significantly.
The effect of KP is clear in the TIL setting: it yields the highest overall accuracy after all five tasks. While both KP and non-KP systems learn successfully the new task at each stage (indicated by the second number in the parentheses), KP preserves performance on earlier tasks better.
Contrary to expectations, jointly trained systems and common-head variants perform worse than TIL. This is likely because the network architecture is optimized for task-incremental learning, and it should be possible to obtain better performance by optimizing the parameters for joint learning. However, we can see the expected decline in accuracy as the number of classes increases. 
We measure the average accuracy per task after the final step for both TIL variants. The results are shown in Figure \ref{fig:barplot}. The KP-model has consistently higher accuracy, particularly for the earlier tasks. In particular, the performance on Task 0 drops from approximately 60\% (Table~\ref{tab:acc}) to 58\% with KP, but to 37\% without it, after training incrementally on all the tasks.
While the gap narrows in later tasks, the KP model maintains stable performance over tasks, in contrast to the vanilla model, which exhibits marked fluctuations and signs of catastrophic forgetting. This suggests that KP favors memory stability, whereas the vanilla model trades it for plasticity, leading to severe forgetting of earlier tasks.

The adopted solution is successful also with other datasets: we experimented with URBANSOUND-8K \cite{salamon_dataset_2014}, with each task made up of 2 classes, for a total of 5 tasks.
The proposed method obtains an accuracy of 84\%, 87\%, 86\%, 85\% and 92\% on the 5 tasks, outperforming the non KP-model by as much as 4\% on the earlier tasks.
We also report three standard continual learning metrics in Table \ref{tab:metrics}. Forgetting Measure (FM) quantifies how much knowledge is lost on previously learned tasks after new tasks are introduced, comparing peak task performance to its final accuracy.  
FM is always non-negative, with lower values indicating better retention. 
The Intransigence Measure (IM) assesses the model's ability to learn new tasks (plasticity), ranging from -1 to 1, where lower values reflect better adaptability.
Based on the values in Table~\ref{tab:metrics}, we observe that the KP model consistently shows lower FM across all steps, confirming its superior ability to retain past knowledge. Meanwhile, both models exhibit similar IM trends, indicating that KP preserves the ability to learn new tasks without significantly sacrificing plasticity.

Finally, Backward Transfer (BWT) measures how learning new tasks affects performance on earlier ones, with positive values indicating beneficial transfer, and negative values indicating forgetting. 
As shown in Table~\ref{tab:metrics} the KP-model achiever BWT values closer to zero, indicating that new tasks interfere less with earlier ones compared to the vanilla model. 
Overall, these results confirm that the proposed  kernel plasticity approach improves performance in incremental learning. Specifically, the KP model demonstrates stronger memory retention and sufficient plasticity, achieving a favorable stability-plasticity trade-off, one of the core objectives of continual learning.
\section{Conclusions}
\label{sec:concl}
This work introduced a biologically inspired solution to catastrophic forgetting by integrating kernel plasticity with Hebbian deep neural networks for incremental audio classification. The proposed neuro-modulation selectively regulates kernel plasticity, enabling the model to preserve past knowledge while adapting to new tasks. Experiments on ESC-50 show clear benefits over a non-modulated version, with overall accuracy improving from 68.7\% to 76.3\%. 
Continual learning metrics confirm that kernel plasticity enhances memory stability without compromising plasticity. These findings highlight the potential of Hebbian learning, complemented by bio-inspired modulation, as an unsupervised alternative to backpropagation in continual learning.
Future work will investigate different datasets and label-free scenarios, and additional biologically motivated mechanisms for unsupervised continual learning. 
\vfill\pagebreak
\bibliographystyle{IEEEtran}
\bibliography{refs}

\end{document}